\documentclass[twocolumn,aps,prb,showpacs,superscriptaddress,groupedaddress]{revtex4}
\usepackage{color}
\usepackage{amsmath}
\usepackage{amssymb}
\usepackage{graphicx}
\usepackage{esint}

\makeatletter

\providecommand{\tabularnewline}{\\}

\@ifundefined{textcolor}{}
{%
 \definecolor{BLACK}{gray}{0}
 \definecolor{WHITE}{gray}{1}
 \definecolor{RED}{rgb}{1,0,0}
 \definecolor{GREEN}{rgb}{0,1,0}
 \definecolor{BLUE}{rgb}{0,0,1}
 \definecolor{CYAN}{cmyk}{1,0,0,0}
 \definecolor{MAGENTA}{cmyk}{0,1,0,0}
 \definecolor{YELLOW}{cmyk}{0,0,1,0}
}

\usepackage{color}
\usepackage{caption}
\usepackage{subcaption}
\usepackage[toc,page]{appendix}\usepackage{setspace}\usepackage{multirow}

\makeatother

\begin{document}

\title{Exciton Quasi-Condensation in One Dimensional Systems}

\date{\today}

\author{Yochai Werman}

\author{Erez Berg}

\affiliation{Weizmann Institute of Science}
\begin{abstract}
A quasi-exciton condensate is a phase characterized by quasi-long
range order of an exciton (electron-hole pair) order parameter. Such
a phase can arise naturally in a system of two parallel oppositely
doped quantum wires, coupled by repulsive Coulomb interactions. We
show that the quasi-exciton condensate phase can be stabilized in
an extended range of parameters, in both spinless and spinful systems.
For spinful electrons, the exciton phase is shown to be distinct from
the usual quasi-long range ordered Wigner crystal phase characterized
by power-law density wave correlations. 
The two phases can be clearly distinguished through their inter-wire
tunneling current-voltage characteristics. 
In the quasi-exciton condensate phase the tunneling conductivity diverges
at low temperatures and voltages, whereas in the Wigner crystal it
is strongly suppressed. 
Both phases are characterized by a divergent Coulomb drag at low temperature.
Finally, metallic carbon nanotubes are considered as a special case
of such a one dimensional setup, and it is shown that exciton condensation
is favorable due to the additional valley degree of freedom. 
\end{abstract}
\maketitle

\section{Introduction}

Excitons are bound states between an electron and a positively charged
hole. Similar to Cooper pairs, which are bound states of two electrons,
excitons are bosons, and may form a condensate. The exciton condensate
phase has been studied extensively, both theoretically and experimentally,
and in the last few years several groups have reported physical signatures
of such a phase in two-dimensional bilayers coupled by Coulomb interactions\cite{MacDonald,Seamons2007,Nandi,DasGupta2010-1}.
Two typical experiments are performed in order to probe exciton condensation
- counter-flow Coulomb drag\cite{DasGupta2010-1,Nandi,Laroche}, in
which the current flow through only one of the layers is suppressed
due to interlayer scattering, and tunneling\cite{Spielman2000}, whereby
the current between the layers is enhanced at low voltages. 

Although long-range excitonic order cannot exist in 1D, an exciton quasi-condensate
phase (corresponding to a power-law decay of exciton correlations)
can occur at zero temperature. Such a phase has definitive signatures
in both tunneling and Coulomb drag experiments. In this paper, we
consider a system of two parallel quantum wires with an opposite sign
of the carriers. We calculate the inter-wire tunneling current-voltage
characteristics in different regimes. We show that a divergence of
the tunneling conductivity at low temperatures and voltages is intimately
linked to exciton quasi-condensation, while a peak in the drag resistance
is not. In addition, we consider the special case of a pair of oppositely
gated parallel carbon nanotubes, 
for which we show that exciton quasi-condensation is particularly
favorable\cite{comment-Min2008}.

For simplicity, we begin with a spinless model (Section II). The
addition of spin in Section III introduces an incompatibility between
tunneling and interwire backscattering, resulting in modified $I-V$
curves corresponding to the case of either tunneling or backscattering
dominated systems. To conclude, we consider the case of carbon nanotubes
(Section IV), which are a natural experimental realization of our
model, and display an additional electronic degree of freedom - the
valley. 

\begin{figure}
\centering \includegraphics[width=2.5in]{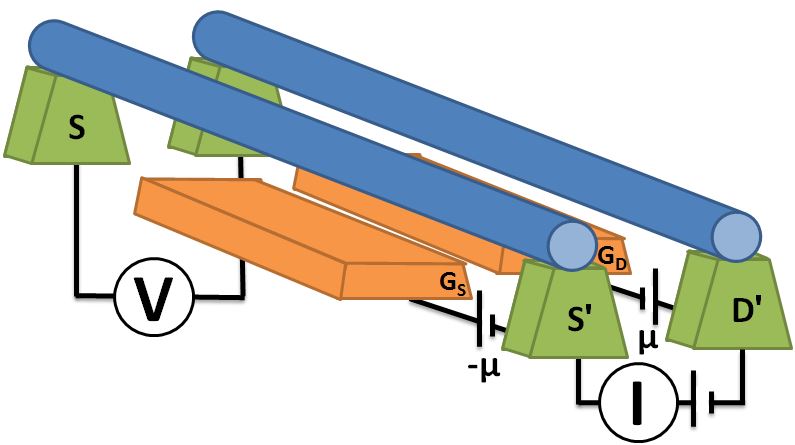}
\caption{Proposed setup of the system. The two wires (in this figure - carbon
nanotubes) are suspended via the green metallic contacts (color online)
above additional gates. The voltage difference between the contacts
and the gates controls the doping of the nanotubes, and we choose
this voltage to be opposite for the two wires ($\mu$ and $-\mu$
in the figure) so that CNT S is electron doped and CNT D hole doped,
with the same density of charge carriers. In a tunneling experiment,
a voltage difference is placed between contacts $S'$ and $D'$, forcing
a tunneling current between the two nanotubes, which is measured.
The resulting voltage difference between the two nanotubes may be
read off the voltmeter placed between $D$ and $S$, which gives $V+2\mu$,
$V$ being the desired quantity.}
\label{fig:Setup1}
\end{figure}

 \begin{figure} \centering
\includegraphics[height=2in,width=2.5in]{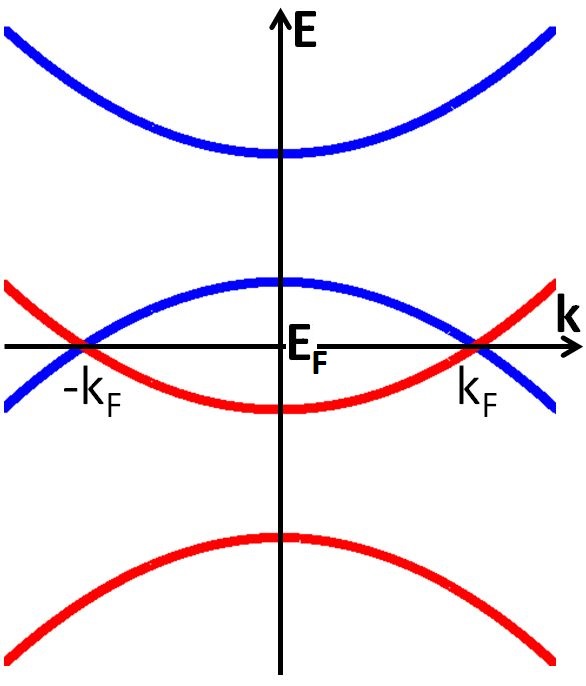} \caption{The dispersion relation for the two wires. The upper, red dispersion (color online) represents the electron-doped wire, while the other is hole doped. The densities of charge carriers
are the same for both wires, and therefore $k_{F}$ is identical. Tunneling between the two wires connects right movers from
one wire to left movers from the other. The spectrum is linearized
around each Fermi point, and the slopes, equal to the Fermi velocity
$v_{F}$, are assumed to be of the same magnitude for simplicity.}
\label{fig:Setup3} 
\end{figure}

\section{Spinless Fermions}

\subsection{Model}

We commence our analysis with spinless fermions. The theoretical system
under consideration is composed of two parallel, infinitely long one
dimensional wires, one of which is doped with electrons and the other
with holes (See Fig.\ref{fig:Setup1}). 
The density of carriers in the two wires is identical, so that the
Fermi momenta are equal, $k_{F1}=k_{F2}$. The system is described
by the Hamiltonian

\begin{equation}
H=H_{\mathrm{kin}}+H_{\mathrm{int}}.
\end{equation}
Here, $H_{\mathrm{kin}}=-i\sum_{j,\eta}v_{F}\int dxc_{j\eta}^{\dagger}(j\eta\partial_{x})c_{j\eta}$,
where $c_{j\eta}^{\dagger}$ is the creation operator for a right/left
moving ($\eta=\pm1$) electron in wire $j=\pm1$; the spectrum has been linearized around $k_F$. $H_{\mathrm{int}}$
is the interaction Hamiltonian, to be specified below. At this stage,
we neglect the inter-wire tunneling. A system of two wires coupled
by four-fermion interactions (but not by single-electron tunneling) has been studied by various authors (see, e.g., Refs.~[\onlinecite{Tsvelik,Stern,Pustilnik}]);
the new element in the present discussion is the fact that the charges in the two wires are of opposite sign.

It is helpful to revert to the bosonized representation of one dimensional
electrons\cite{Giamarchi}, where $c_{i\eta}^{\dagger}\sim e^{i\left(\eta\phi_{i}-\theta_{i}\right)}$.
The bosonic field $\phi_{i}$ is related to the electronic charge
density fluctuations by $\sum_{\eta}c_{i\eta}^{\dagger}c_{i\eta}=\partial_{x}\phi_{i}/\pi$.
$\theta_{i}$ is similarly related to the electronic current fluctuations.
In terms of the bosonic variables, the small momentum (forward scattering)
part of the interaction becomes quadratic. The Hamiltonian takes the
form $H=H_{0}+H_{1}$, where the quadratic part of the Hamiltonian
results in the Tomonaga-Luttinger form 
\begin{eqnarray}
H_{0}=\sum_{\lambda=\pm}\frac{u_{\lambda}}{2\pi}\int\mbox{dx}\left[K_{\lambda}\left(\partial_{x}\theta_{\lambda}\right)^{2}+\frac{1}{K_{\lambda}}\left(\partial_{x}\phi_{\lambda}\right)^{2}\right],
\end{eqnarray}
where we have introduced the fields $\phi_{\pm}=\frac{1}{\sqrt{2}}\left(\phi_{1}\pm\phi_{2}\right)$,
and the same for $\theta_{\pm}$. $u_{\pm}$ are the velocities of
the plasmons, and $K_{\pm}$ are the corresponding Luttinger parameters, given by $K_{\pm}=\frac{K}{\sqrt{1\pm UK}}$,
where $K$ is the Luttinger parameter of an individual wire (including
the effect of the intra-wire interactions). $U=V_{q=0}/2v_F$ is
the dimensionless inter-wire forward scattering strength ($V_{q}$
is the Fourier transform of the inter-wire density-density interaction).

In addition to the forward scattering term, which is quadratic in
the bosonic fields, backscattering between the wires is also possible,
and gives rise to the term

\begin{align}
H_{BS} & =V_{q=2k_{F}}\int\mathrm{dx}\,c_{1R}^{\dagger}c^{\vphantom{\dagger}}_{1L}c_{-1R}^{\dagger}c^{\vphantom{\dagger}}_{-1L}+\mathrm{H.c.}\nonumber \\
 & =-\left|V_{BS}\right|\int\mbox{dx}\,\cos\left(2\sqrt{2}\phi_{+}\right),\label{eq:H_BS}
\end{align}
where $V_{BS}\propto V_{q=2k_{F}}$ is the strength of the backscattering
interactions. Note the minus sign in Eq. (\ref{eq:H_BS}), which arises
from the commutation relations between the $\phi$ and $\theta$ fields, see Appendix A.
If only small momentum scattering is present ($V_{BS}=0$), the two
modes are massless, and the correlations of physical observables decay
as power laws for any interaction strength. 
The operators with the most slowly decaying correlations are the $2k_{F}$
component of the density, giving rise to the density-density correlation
function decaying as $\left\langle c_{j,+}^{\dagger}(x)c^{\vphantom{\dagger}}_{j,-}(x)c_{j,-}^{\dagger}(0)c{\vphantom{\dagger}}_{j,+}(0)\right\rangle \sim1/x^{K_{+}+K_{-}}$,
and exciton (particle-hole pair) correlations, which satisfy $\left\langle c_{1,+}^{\dagger}(x)c^{\vphantom{\dagger}}_{2,-}(x)c_{2,-}^{\dagger}(0)c{\vphantom{\dagger}}_{1,+}(0)\right\rangle \sim1/x^{K_{+}+1/K_{-}}$.

In the presence of backscattering ($V_{BS}\ne0$), and if backscattering
is relevant (which is the case for $K_{+}<1$), a gap $\Delta_{BS}\propto V_{BS}^{1/(2-2K_{+})}$
opens in the spectrum of total charge fluctuations. The partially
gapped phase has enhanced Wigner crystalline (density-density) correlations.
As we will show, in the spinless case this phase also displays enhanced
excitonic correlations. 

\subsection{Inter-wire Tunneling}

Tunneling current measurements are a sensitive experimental method
by which to probe exciton correlations. Tunneling from an external
lead into an interacting one dimensional system is normally suppressed\cite{Aristov1,Bockarth,Auslaender,Jompol};
this is a result of the strong correlations between electrons in a
Luttinger liquid, which resist the entrance of an external, uncorrelated
particle. In our system of two coupled wires, on the other hand, exciton
correlations between the two wires tend to enhance the tunnelling
current, since particles in one wire are aligned with holes in the
other.

In an experiment by Spielman et al. \cite{Spielman2000}, signatures
of exciton condensation have been detected in a two dimensional bilayer
subject to a high magnetic field. A sharp peak in the differential
inter-layer tunneling conductivity, $\sigma(V)=\frac{dI_{t}}{dV}$ (where $I_{t}$
is the tunneling current density and $V$ the interlayer voltage)
at zero bias has been interpreted as a signature of such long-range
order \cite{Rosenow}. Here, we present theoretical predictions regarding the tunneling
current-voltage relations in the one dimensional equivalent of the
system studied in Ref. [\onlinecite{Spielman2000}].

In order to allow for a tunneling current, weak interwire hopping
is added to the Hamiltonian: 
\begin{eqnarray}
H_{\mathrm{tun}}=- & \tilde{t}_{\perp} & \sum_{i\eta}\int\mbox{dx}c_{i\eta}^{\dagger}(x)c_{-i-\eta}(x)=\notag\\
- & t_{\perp} & \int\mbox{dx}\cos\left(\sqrt{2}\phi_{+}\right)\sin\left(\sqrt{2}\theta_{-}\right),
\end{eqnarray}
where $\tilde{t}_{\perp}$ is
the inter-wire tunnellng amplitude, and $t_{\perp}\propto\tilde{t}_{\perp}$
(the proportionality constant being $1/(2\pi a)$ in a naive continuum
limit, where $a$ is the short distance cutoff). The RG equations
for the two non-quadratic terms, the interwire backscattering (3)
and tunneling, are: 
\begin{eqnarray}
 &  & \frac{dV_{BS}}{ds}=\left[2-2K_{+}\right]V_{BS}\\
 &  & \frac{dt_{\perp}}{ds}=\left[2-\frac{1}{2}\left(K_{+}+\frac{1}{K_{-}}\right)\right]t_{\perp}\notag
\end{eqnarray}
Here $s$ is the momentum scaling parameter, $\Lambda(s)=\Lambda_{0}e^{-s}$.

Tunneling is relevant for $K_{+}+K_{-}^{-1}<4$, which corresponds
to a wide range of physical parameters. In this regime, a gap $\Delta_{t}\propto t_{\perp}^{1/(2-\frac{1}{2}(K_{+}+K_{-}^{-1}))}$
is opened for fluctuations of the total density and relative charge.
In addition, when backscattering is relevant, which is the case for
$K_{+}<1$, the fluctuations of the total density are further suppressed,
renormalizing the corresponding Luttinger parameter to zero. We assume
that the backscattering gap is larger than the tunneling energy scale
in the rest of the section.

For energies above $\Delta_{t}$, or when tunneling is irrelevant,
a linear response calculation for the current is applicable. The tunneling
current is approximated by \cite{Mahan} 
\begin{eqnarray}
I_{t}(V)=2|t_{\perp}|^{2}\mathcal{I}m\{G_{A}^{ret}(q=0,\omega=-eV)\},\label{eq:I_t}
\end{eqnarray}
Where $G_{A}^{ret}(q,t)=-i\Theta(t)\left\langle \left[A(q,t),A^{\dagger}(-q,0)\right]\right\rangle $
and $A(x,t)=c_{1}^{\dagger}(x,t)c_{-1}(x,t)$. The linear response
calculations result in power laws in the voltage, with the exponents
governed by the Luttinger parameters, corresponding to the fact that
current dissipates through the excitation of plasmons in the Luttinger
liquid. The exponents distinguish between regimes where different
sectors are locked; for $\Delta_{BS}\ll eV$, we get $I_{t}\sim V^{K_{+}+1/K_{-}-2}$,
while for $\Delta_{t}\ll eV\ll\Delta_{BS}$ 
 fluctuations in the total density are suppressed, and $I_{t}\sim V^{1/K_{-}-2}$.

If tunneling is relevant, the low voltage behavior ($eV\ll\Delta_{t}$)
is dramatically different. The relative phase field $\theta_{-}$
is locked, signifying that charge may fluctuate freely between the
two wires, without exciting plasmons. In this regime, the perturbative
expression for the current, Eq. (\ref{eq:I_t}), breaks down.

In order to analyze this case, we use a generalization of the ``tilted
washboard'' model for Josephson junctions \cite{Tinkham}. Imagine driving a small current density $\mathcal{J}$
between the two wires. The system is described by following effective
Hamiltonian: 
\begin{equation}
H_{\mathrm{eff}}=H_{0}-\int dx[J\sin(\sqrt{2}\theta_{-})-\frac{\sqrt{2}}{e}\mathcal{J}\theta_{-}],
\end{equation}
where $J=t_{\perp}\langle\cos(\sqrt{2}\phi_{+})\rangle$. (We assume
that the field $\phi_{+}$ is pinned to zero by $H_{\mathrm{BS}}$.)
According to the Josephson relation, the inter-wire current density
operator is given by $eJ\sin(\sqrt{2}\theta_{-})$, and the voltage
is given by $(\sqrt{2}/e)d\theta_{-}/dt$. Dissipation occurs 
by the creation of soliton-antisoliton pairs that dissociate and induce
phase slips of $2\pi$ in $\sqrt{2}\theta_{-}$. The voltage is given
by 
\begin{eqnarray}
V=\frac{2\pi}{e}\Gamma,
\end{eqnarray}
with $\Gamma$ the dissociation
rate. These propagating soliton-antisoliton pairs correspond to macroscopic
quantum tunneling between consecutive minima of the sine-Gordon potential,
$\cos(\sqrt{2}\theta_{-})$, whose degeneracy is broken by the tunneling
current. We have calculated the rate of macroscopic quantum tunneling
using the instanton method (see Appendix B for the details
of the calculation). The result is $\Gamma\propto e^{-\frac{\alpha}{\mathcal{J}}}$,
with $\alpha\approx 2.5e\sqrt{\frac{u_-K_-}{2\pi}J}$. This leads to the highly singular current-voltage
relationship 
\begin{eqnarray}
I(V)\propto-\frac{1}{\log\left(V\right)}.
\end{eqnarray}
Here $I$ is the tunneling current and $V$ the interwire voltage. 

Lastly, for high enough temperatures such that $T\gtrsim eV$, the
low voltage behavior is Ohmic: $I\propto f(T)V$, where $f(T)=T^{-(3-K_+-1/K_-)}$; the retarded Green's function at finite temperatures is given by\cite{Giamarchi}
\begin{eqnarray}
&&G^{ret}(q=0,\omega)\propto\\
&&T^{\alpha-2}B\left(-i\frac{\omega}{4\pi T}+\frac{\alpha}{4},1-\frac{\alpha}{2}\right)^2\notag,
\end{eqnarray}
with $\alpha=K_++1/K_-$, and $B$ the $\beta$-function. Expanding to first order in $\omega$, this gives an imaginary part proportional to $T^{\alpha-3}$.

\begin{figure}
\centering \includegraphics[width=3.5in]{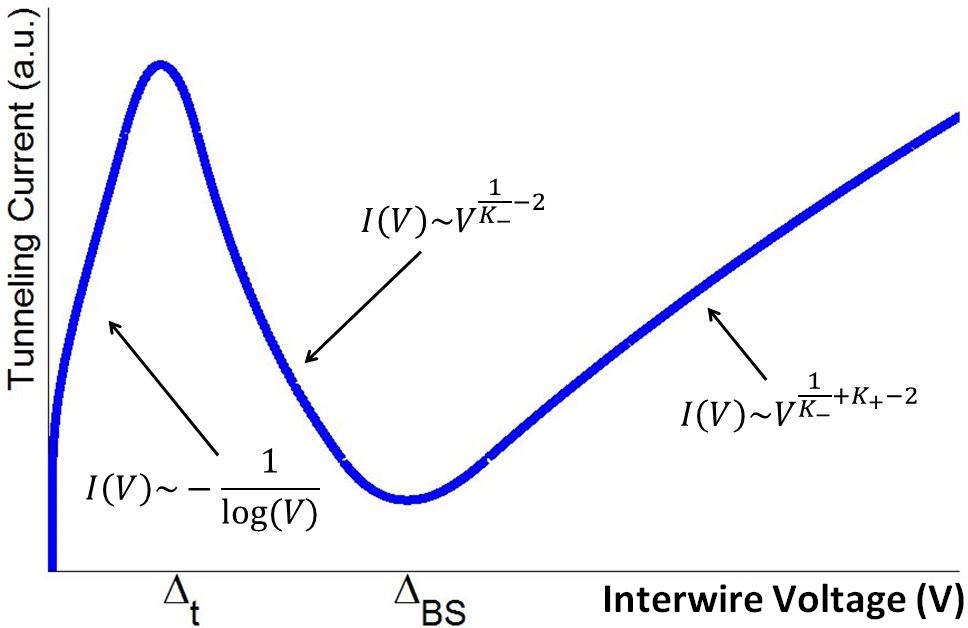} \caption{Current-voltage characteristics for spinless fermions, in the case of
relevant interwire tunneling. The tunneling current obeys the singular
relationhsip $I(V)\propto-1/\log(V)$ for voltages lower than the
tunneling gap, while at higher voltages it reverts to a power law
in the voltage, with the exponent increasing as the difference is
amplified. For a large set of values of $K_{+},$ $K_{-}$ quite accessible
experimentally, the exponent may change between a negative and a positive
value, as shown in this figure, corresponding to a negative differential
conductance for intermediate voltages. This is because in the regime
$\Delta_{t}\ll V \ll\Delta_{BS}$, the total sector is locked by the
backscattering gap.}

\label{fig:I(V)} 
\end{figure}

Our results for spinless electrons are summarized in Table \ref{table:results},
and a typical I(V) curve is displayed in Fig.\ref{fig:I(V)}.

\begin{table*}
\begin{tabular}{|c|c|c|c||c|c|}
\hline 
\multicolumn{1}{|c|}{} & \multicolumn{3}{c||}{$T=0$} & \multicolumn{2}{c|}{$T\gg\Delta_{t},\Delta_{BS}$}\tabularnewline
\hline 
\multicolumn{1}{|c|}{} & $\Delta_{t}\gg V$  & $\Delta_{BS}\gg V\gg\Delta_{t}$  & $V\gg\Delta_{BS}$  & $V\ll T$  & $V\gg T$ \tabularnewline
\hline 
\multicolumn{1}{|c|}{$\mbox{Tunneling relevant}$} & $I(V)\propto\frac{-1}{\log(V)}$  & $I(V)\propto V^{1/K_{-}-2}$  & $I(V)\propto V^{K_{+}+1/K_{-}-2}$  & \multirow{2}{*}{$I(V)\propto f(T)V$}  & \multirow{2}{*}{$I(V)\propto V^{K_{+}+1/K_{-}-2}$}\tabularnewline
\cline{1-4} 
\multicolumn{1}{|c|}{\mbox{Tunn. irrelevant}} & \multicolumn{2}{c|}{$I(V)\propto V^{1/K_{-}-2}(\Delta_{t}=0)$} & $I(V)\propto V^{K_{+}+1/K_{-}-2}$  & \multicolumn{1}{c|}{} & \multicolumn{1}{c|}{}\tabularnewline
\hline 
\end{tabular}\caption{The tunneling current dependence on interwire voltage $V$ for spinless
fermions. When $K_{+}+1/K_{-}<4$, tunneling is a relevant. For voltages
below the tunneling gap $\Delta_{t}$, the tunneling I-V relationship
is a singular function, resulting in a diverging tunneling resistivity.
For voltages above the gap, and when tunneling is irrelevnat, the
I-V curves obey power laws, as expected in one dimensional systems,
with the exponents disinguishing between the regime in which the total
density is locked by the relevant backscattering process to that in
which it is free. The high temperature system displays ohmic behavior
at low voltages.}

\label{table:results} 
\end{table*}

The low voltage tunneling current has clear signatures of exciton
correlations. When the inter-wire tunneling is relevant, 
the zero bias conductivity
$\lim_{V\rightarrow0}\frac{dI}{dV}$ diverges at low temperatures.
Furthermore, in the regime in which excitonic correlations are strong,
the $I-V$ curve may exhibit negative differential conductance for
a finite range of voltages, $\Delta_{t}\ll V\ll\Delta_{BS}$, which
may also be interpreted as a mark of exciton quasi-condensation.

\section{Spinful electrons}

We now consider the case of spinful electrons. The quadratic part of the 
Hamiltonian is 
\begin{eqnarray}
H_{0}=\sum_{\mu,\lambda=\pm}\frac{u_{\mu\lambda}}{2\pi}\int\mbox{dx}\left[K_{\mu\lambda}\left(\partial_{x}\theta_{\mu\lambda}\right)^{2}+\frac{1}{K_{\mu\lambda}}\left(\partial_{x}\phi_{\mu\lambda}\right)^{2}\right]\notag\\
\end{eqnarray}
where, following standard notation, $\phi_{\mu\pm}=\frac{1}{\sqrt{2}}\left(\phi_{1\mu}\pm\phi_{2\mu}\right)$,
$\mu=\rho,\sigma$, where $\phi_{i,\rho/\sigma}=\frac{1}{\sqrt{2}}(\phi_{i\uparrow}\pm\phi_{i\downarrow})$.
$\theta_{\rho,\pm}$, $\theta_{\sigma,\pm}$ are defined in a similar
fashion. $u_{\sigma\pm}$, $u_{\rho\pm}$ are the velocities of the
spin and charge plasmons, and $K_{\sigma\pm}$, $K_{\rho\pm}$ are
the corresponding Luttinger parameters. Assuming that there are only
density-density inter-wire interactions, spin rotation invariance
requires $K_{\sigma\pm}=1$. The charge Luttinger parameters are $K_{\rho\pm}=\frac{K_{\rho}}{\sqrt{1\pm UK_{\rho}}}$,
where $K_{\rho}$ is the charge Luttinger parameter of an individual
wire.

In the spinful model, the $2k_{F}$ interwire backscattering is adverse
to exciton quasi-condensation; it takes the form 
\begin{eqnarray}
 && H_{BS}=\\
&&-|V_{BS}|\int\mbox{dx}\cos\left(2\phi_{\rho+}\right)\left[\cos\left(2\phi_{\sigma+}\right)+\cos\left(2\phi_{\sigma-}\right)\right],\notag
\end{eqnarray}
while the tunneling of spinful electrons is described by (see Appendix A)
\begin{eqnarray}
H_\mathrm{tun} &=& t_{\perp}\int\mbox{dx}[\cos\left(\phi_{\rho+}\right)\cos\left(\phi_{\sigma+}\right)\cos\left(\theta_{\sigma-}\right)\sin\left(\theta_{\rho-}\right)\notag\\
&&-\sin\left(\phi_{\rho+}\right)\sin\left(\phi_{\sigma+}\right)\sin\left(\theta_{\sigma-}\right)\cos\left(\theta_{\rho-}\right)]
\end{eqnarray}
It can be seen that the backscattering term tends to lock the field $\phi_{\sigma -}$ (that describes relative spin fluctuations), and therefore it suppressed the transfer of electrons between the wires. Thus the tunnelling and backscattering terms compete with each other in this case. The scaling equations for these two terms are 

\begin{eqnarray}
 & \frac{dV_{BS}}{ds}=\left[1-K_{\rho+}\right] V_{BS},  \nonumber
  \\ 
 & \frac{dt_{\perp}}{ds}=\left[\frac{3}{2}-\frac{1}{4}\left(K_{\rho+}+\frac{1}{K_{\rho-}}\right)\right]t_{\perp}.
\end{eqnarray}

The spinful model leads to two distinct phases, depending on the various
Luttinger parameters and initial amplitudes of the non quadratic terms:
\begin{enumerate}
\item Tunneling is the dominant interaction: tunneling is more relevant
than the $2k_{F}$ backscattering term for $K_{\rho}>\frac{1}{3}-\frac{1}{36}U$, 
and tends to open a gap $\Delta_{t}\propto t_{\perp}^{1/(3/2-1/4(K_{\rho+}+1/K_{\rho-}))}$
for fluctuations of the density in the total sectors and of the phase
in the relative sectors. 
\item Backscattering is dominant: In this regime, the $2k_{F}$ backscattering term opens
a gap $\Delta_{BS}\propto |V_{BS}|^{1/(1-K_{\rho+})}$ for relative spin fluctuations, which suppresses tunneling up
to $\Delta_{BS}$. This phase is characterized by a spin gap and quasi-long ranged 
charge density waves correlations at wavevector $2k_F$. We denote it
as a Wigner crystal. 
\end{enumerate}

\begin{figure*}
      \centering
        \begin{subfigure}{.45\textwidth}
        		\centering
                \includegraphics[height=2in]{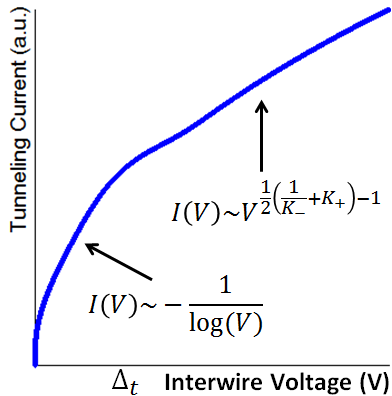}
                \caption{Tunneling dominant regime}
                \label{fig:relevant}
        \end{subfigure} %
        \begin{subfigure}{.45\textwidth}
        		\centering
                \includegraphics[height=2in]{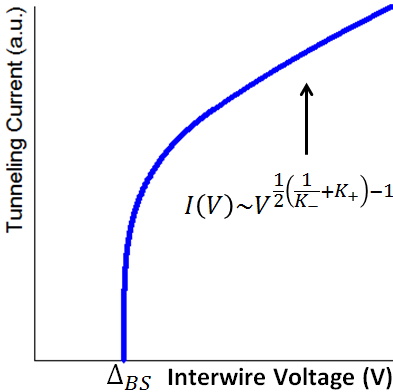}
                \caption{Backscattering dominant regime}
                \label{fig:irrelevant}
        \end{subfigure}
        
        \caption{$I-V$ curve for spinful electrons in the regimes where tunneling or backscattering is dominant. In the tunneling dominant regime, the low voltage current obeys the singular relationship $I(V)\propto\frac{-1}{\log(V)}$,
        as the current dissipates through the macroscopic tunneling mechanism
        described in Section II. At high voltages, the current is a powerlaw
        in the voltage, $I(V)\propto V^{\frac{1}{2}\left(K_{\rho+}+K_{\rho-}^{-1}\right)-1}$.
        The presence of noninteracting spin sectors ($K_{\sigma}=1$), and interwire forward scattering $U$, may
        decrease the exponent to negative values, resulting in a negative
        differential conductance. In the regime where backscattering is dominant, the
        interaction opens a gap $\Delta_{BS}$ for relative spin fluctuations
        which suppresses tunneling up to that scale. At $V\gg\Delta_{BS}$, the current obeys
        the powerlaw $I(V)\propto V^{\frac{1}{2}\left(K_{\rho+}+K_{\rho-}^{-1}\right)-1}$. }\label{fig:Spinful}
\end{figure*}

The zero temperature tunneling current-voltage characteristics of the two phases are shown in Figures ~\ref{fig:relevant} and~\ref{fig:irrelevant}. When the voltage is greater than the gap (either $\Delta_t$ or $\Delta_{BS}$), the current follows a powerlaw dependence,  $I(V)\propto V^{\frac{1}{2}(K_{\rho+} + {K^{-1}_{\rho-}} - 1) }$. The exponent can have either sign, depending on the strength of the inter- vs. intra-wire interactions. In the backscattering dominated phase, the current goes abruptly to zero when the voltage is below $\Delta_{BS}$. In contrast, in the tunnelling dominated phase, at small voltages the current has a logarithmic dependence on voltage, $I(V) \propto -1/\log(V/\Delta_t)$, as in the spinless case. The tunneling conductivity,
$\sigma=\lim_{V\rightarrow0}\frac{dI}{dV}$, diverges at low voltages.


If the intra-wire interactions are strong, additional backscattering terms arise. In this case, the density correlations in each wire become strongly peaked at a wave vector of $4k_F$, and the amplitude of the $4k_F$ ultimately becomes stronger than the $2k_F$ component~\cite{White2002}. The coupling of the $4k_F$ Fourier components of the density in the two wires leads to an additional backscattering term of the form
\begin{eqnarray}
 && H_{BS,4k_F}=-|V_{BS,4k_F}|\int\mbox{dx}\cos\left(4\phi_{\rho+}\right).
\end{eqnarray}
This term, while less relevant than $H_{BS}$, has a large amplitude in the strongly interacting limit. In this case, a gap in the $\rho +$ sector can open at an energy larger than $\Delta_t$, $\Delta_{BS}$. In the tunnelling dominated phase, this will result in an $I-V$ curve more similar to the one shown in Fig.~\ref{fig:I(V)}, with two characteristic energy scales: $\Delta_{BS,4k_F}$ at which the $\rho+$ sector becomes gapped, and $\Delta_t$. We will discuss the effects of the $4k_F$ backscattering term in more detail in a future publication. 



Before concluding, let us mention the signatures of the two phases in the drag resistance between the two wires. Although $2k_F$ backscattering suppresses excitonic correlations
due to the locking of the relative spin density, the drag resistance diverges in the backscattering dominated phase
at low temperatures \cite{Stern,Pereira,Teber,Pustilnik,Flensberg,Nazarov1,Ponomarenko,Fiete},
since the $\phi_{\rho+}$ is locked. In this respect, a divergent drag resistance does not signify enhanced excitonic correlations in our system. The tunnelling dominated phase is insulating at temperatures below $\Delta_t$, so measuring drag resistance is impossible unless a large bias voltage $V>\Delta_t$ is applied to one of the wires.


\vspace{5mm}

\section{Carbon nanotubes}

An obvious experimental realization of the system considered in this work is a double carbon nanotube setup. The two metallic nanotubes can be brought closely together
and gated independently. 
The linearity of the spectrum insures particle-hole
symmetry, which favors exciton formation, as such a symmetry results
in a nested Fermi surface towards the creation of particle-hole excitations
with zero momentum. Finally, carbon nanotubes may be fabricated with
an exceptional purity, rendering our neglect of disorder in the previous
analysis tenable.

In addition to spin, metallic carbon nanotubes have a valley degree of freedom \cite{Dresselhaus}. Applying a magnetic field along the axis of the nanotubes lifts the valley degeneracy, as well as the spin degeneracy; one can design systems in which either the valley, the spin, or both are quenched (i.e., only a electrons of a single spin or valley flavour cross the Fermi energy) \cite{Ilani}. In these cases, the analysis of the previous two sections goes through without modification. Here, we comment briefly on the valley degenerate case (no magnetic field). 

Denoting the valley label by $\nu=\pm1$, we
define the boson fields $\phi_{\mu\pm t}=\frac{1}{\sqrt{2}}\left(\phi_{\mu\pm\nu=1}+\phi_{\mu\pm\nu=-1}\right)$
and $\phi_{\mu\pm r}=\frac{1}{\sqrt{2}}\left(\phi_{\mu\pm\nu=1}-\phi_{\mu\pm\nu=-1}\right)$,
with $\mu=\rho,\sigma$ as before. Intervalley scattering is very
much suppressed in carbon nanotubes~\cite{Kane1997}, and therefore we assume that $K_{\rho+ r}=K_{\sigma+ t}=K_{\sigma+ r}=K_{\rho- r}=K_{\sigma -t}=K_{\sigma- r}=1$. The interaction strength is encoded in the Luttinger parameters $K_{\rho \pm t} =  \frac{K_{\rho t}}{\sqrt{1\pm UK_{\rho t}}}$, where $K_{\rho t}$ is the Luttinger parameter for the total charge for a single wire, and $U$ the interwire forward scattering potential.

As in the spinful valleyless case, the interwire backscattering term (16) tends to lock both total ($+$) and relative ($-$) modes, and thus competes with the tunneling term (17). The backscattering term is written as
 \begin{widetext}
\begin{eqnarray}
H_{BS}=&-&|V_{BS}|\times[\cos\left(\sqrt{2}\phi_{+\rho t}\right)\cos\left(\sqrt{2}\phi_{+\rho r}\right)\left(\cos\left(\sqrt{2}\phi_{+\sigma t}\right)\cos\left(\sqrt{2}\phi_{+\sigma r}\right)+\cos\left(\sqrt{2}\phi_{-\sigma t}\right)\cos\left(\sqrt{2}\phi_{-\sigma r}\right)\right)\notag\\
&+&\cos\left(\sqrt{2}\phi_{+\rho t}\right)\cos\left(\sqrt{2}\phi_{-\rho r}\right)\left(\cos\left(\sqrt{2}\phi_{+\sigma t}\right)\cos\left(\sqrt{2}\phi_{-\sigma r}\right)+\cos\left(\sqrt{2}\phi_{-\sigma t}\right)\cos\left(\sqrt{2}\phi_{+\sigma r}\right)\right)],
\end{eqnarray}
while the inter-wire tunnelling term has the form
\begin{eqnarray}
H_{tun}=t_{\perp}\sin\left(\frac{\phi_{+\rho t}}{\sqrt{2}}\right)\cos\left(\frac{\phi_{+\rho r}}{\sqrt{2}}\right)\cos\left(\frac{\phi_{+\sigma t}}{\sqrt{2}}\right)\cos\left(\frac{\phi_{+\sigma r}}{\sqrt{2}}\right)\cos\left(\frac{\theta_{-\rho t}}{\sqrt{2}}\right)\cos\left(\frac{\theta_{-\rho r}}{\sqrt{2}}\right)\cos\left(\frac{\theta_{-\sigma t}}{\sqrt{2}}\right)\cos\left(\frac{\theta_{-\sigma r}}{\sqrt{2}}\right)\notag\\
+\mbox{...,}
\end{eqnarray}
\end{widetext}
where the ellipsis stands for terms in which an odd number of cosines is exchanged with sines.
The lowest order $\beta$-functions of the two terms are given by
\begin{eqnarray}
& \frac{dV_{BS}}{ds}=\frac{1}{2}\left[1-K_{+\rho t}\right] V_{BS},  \nonumber
  \\ 
 & \frac{dt_{\perp}}{ds}=\left[\frac{5}{4}-\frac{1}{8}\left(K_{\rho+}+\frac{1}{K_{\rho-}}\right)\right]t_{\perp}.
\end{eqnarray}
The addition of internal degrees of freedom, such as valley, enhances the possibility for
exciton condensation, as was already seen in Section III; the phase where
the tunneling conductivity is the most divergent corresponds to intrawire
Luttinger parameters obeying $K_\rho>0.15-0.01U$ for small $U$. This regime is accessible in experiments; 
the experimental estimates for the Luttinger parameter of single walled nanotubes\cite{Bockarth,Yao1999} are in the range $K_\rho \sim 0.2-0.3$, resulting in a more divergent tunneling operator, and thus the tunneling dominated phase may be reached. In this phase, the low voltage tunneling current will obey the familiar $I(V)\propto\frac{-1}{\log(V)}$ law, while for voltages much higher than any emergent gap in the system, the current will again be a power law in the voltage, with the exponent modified by the additional noninteracting sectors, resulting in $I(V)\propto V^{\frac{1}{4}\left(K_{\rho+t}+K_{\rho-t}^{-1}\right)-\frac{1}{2}}$.

\section{Conclusion}

In conclusion, we consider a system of oppositely doped wires in the
limit of strong forward scattering and weak interwire backscattering
and tunneling. For strong enough interwire interactions, the system
is susceptible to excitonic quasi-long range order. We further discuss
the tunneling current-voltage characteristics, and show that there
are three different regimes, depending on the relative relevance and
magnitude of the tunneling process and the two types of interwire
backscattering. When the tunneling process is dominant, corresponding
to a phase with strong excitonic correlations, the zero bias conductivity
diverges, making tunneling a sensitive probe of interwire phase coherence.
On the other hand, we argue that the drag resistance will diverge
both in the excitonic regime and in the Wigner crystal, where the
phases of the two wires are independent. Lastly, we examine carbon
nanotubes, and note that they are exceptionally suited for the detection
of excitons.

\acknowledgements{We thank S. Ilani and B. Rosenow for useful discussions. This work was supported by the Israel Science Foundation, the Minerva fund and a Marie Curie Career Integration Grant (CIG).}

\appendix

\section{Sign Issues}
In bosonization theory, electrons of the same species obey fermion statistics due to the commutation relations of the bosons $\phi$ and $\theta$, while the anticommutation of different species must be put in by hand\cite{Giamarchi}. One way to achieve this is to impose the following relationship between $\theta$s of different electrons\cite{Delft}:
\begin{eqnarray}
[\theta_i(x),\theta_j(x')]=i\pi\epsilon_{ij},
\end{eqnarray}
where $\epsilon_{ij}$ is the antisymmetric tensor.
\subsection{Interwire backscattering}
One consequence of the anticommutation of fermions is the sign of the backscattering term. Consider the operator 
\begin{eqnarray}
&&c_{1R}^\dagger(x) c_{1L}(x) \propto \\
&&\exp\left(i\left(\phi_1(x)-\theta_1(x)\right)\right)\exp\left(-i\left(-\phi_1(x)-\theta_1(x)\right)\right)\notag\\
&&=\exp\left(2i\phi_1(x)\right)e^{\frac{1}{2}[\phi_1(x)-\theta_1(x),-\phi_1(x)-\theta_1(x)]},\notag
\end{eqnarray}
where the second term on the last line is a consequence of the Baker-Hausdorff formula. The commutator can be evaluated easily using the convention
\begin{eqnarray}
[\phi(x),\theta(x')]=i\pi \Theta(x-x')
\end{eqnarray}
along with the point-splitting technique\cite{Giamarchi}; it results in
\begin{eqnarray}
c_{1R}^\dagger(x) c_{1L}(x) \propto -i\exp\left(2i\phi_1(x)\right).
\end{eqnarray}
Therefore, 
\begin{eqnarray}
&&c_{1R}^\dagger(x) c_{1L}(x) c_{2R}^\dagger(x) c_{2L}(x)\propto\\
&&i\exp\left(2i\phi_1(x)\right)\times i\exp\left(2i\phi_2(x)\right)=\notag\\
&&-\exp\left(2\sqrt{2}i\phi_+(x)\right)\notag,
\end{eqnarray}
incurring a minus sign that is absent for wires doped equally with the same kind of charge carrier, where backscattering is of the form $c_{1R}^\dagger c_{1L}c_{2L}^\dagger c_{2R}$.

\subsection{Tunneling}
The form of the tunneling term is also affected by the fermionic phase:
\begin{eqnarray}
c_{1R}^\dagger c_{2L}&\propto& e^{i(\phi_1+\phi_2-\theta_1+\theta_2)}e^{\frac{1}{2}[\theta_1,\theta_2]}\notag\\
&=&ie^{i(\phi_1+\phi_2-\theta_1+\theta_2)},
\end{eqnarray}
with equivalent phases for the other three components of the tunneling interaction, which result in the form given by equation (4).
The addition of spin generates terms of the form
\begin{eqnarray}
c_{1R\uparrow}^\dagger c_{2L\downarrow}=e^{i(\phi_{\rho+}-\theta_{\rho-}+\phi_{\sigma+}-\theta_{\sigma-})}e^{\frac{1}{4}([\theta_{1\rho},\theta_{2\rho}]+[\theta_{1\sigma},\theta_{2\sigma}])},\notag\\
\end{eqnarray}
resulting again in a factor $i$, and leading to equation (13).

\section{Instanton calculation}
The real time action governing the evolution of the relative phase is given by
\begin{eqnarray}
S&=&\frac{u_-K_-}{2\pi}\int{dxdt}\left(\frac{1}{u_-^2}\left(\partial_t\theta_-\right)^2-\left(\partial_x\theta_-\right)^2\right)\notag\\
&+&\int{dxdt}\left[J\sin\left(\sqrt{2}\theta_-\right)+\frac{\sqrt{2}}{e}\mathcal{J}\theta_-\right],
\end{eqnarray}
where, as in section II, $J=t_\perp\left\langle\cos(\sqrt{2}\phi_+)\right\rangle$ and $\mathcal{J}$ is the current density driven between the wires, which breaks the symmetry between the minima of the sine-Gordon potential.

The decay rate of the metastable state $\sqrt{2}\theta_0=\frac{\pi}{2}$ can be calculated from the Green's function
$G(T)=\left\langle \theta_ 0\right| e^{iHT} \left| \theta_0\right\rangle,$ which after analytic continuation becomes
\begin{eqnarray}
G(\tau)=\left\langle \theta_ 0\right| e^{H\tau} \left| \theta_0\right\rangle=\int\mathcal{D}[\theta(\tau')]e^{-S_E[\theta(\tau')]},
\end{eqnarray}
where the Euclidean action $S_E$ is given by
\begin{eqnarray}
S&=&\frac{u_-K_-}{2\pi}\int{dxd\tau}\left(\frac{1}{u_-^2}\left(\partial_\tau\theta_-\right)^2+\left(\partial_x\theta_-\right)^2\right)\notag\\
&-&\int{dxd\tau}\left[J\sin\left(\sqrt{2}\theta_-\right)+\frac{\sqrt{2}}{e}\mathcal{J}\theta_-\right].
\end{eqnarray}

We evaluate this path integral by a saddle point approximation, corresponding to field configurations $\{\theta_-(x,\tau)\}$ which satisfy the Euler-Lagrange equations
\begin{eqnarray}
\partial_r^2\theta+\frac{1}{r}\partial_r\theta=-\frac{\sqrt{2}\pi}{uK}\left[J\cos(\sqrt{2}\theta)+\frac{\mathcal{J}}{e}\theta\right],
\end{eqnarray}
where we have defined $r=\sqrt{x^2+u^2\tau^2}$.

For $\mathcal{J}\ll eJ$, meaning that the potential minimum asymmetry imposed by the current $\mathcal{J}$ is small, it is possible to use the thin wall approximation\cite{Altland}; in this case, it is assumed that the configurations which minimize the action are described by a domain wall of thickness $\Delta r$ which is positioned at $r_0\gg\Delta r$, which separates regions of homogenous configurations of $\theta$ - $\theta(r<r_0)=\frac{\pi}{2}, \theta(r>r_0)=\frac{5\pi}{2}$. In this case, the second term in the left hand side of Eq. B4 is much smaller than the first, and the Euler-Lagrange equations are
\begin{eqnarray}
\partial_r^2\theta\approx-\frac{2\sqrt{2}\pi}{uK}\left[J\cos(\sqrt{2}\theta)+\frac{\mathcal{J}}{e}\theta\right].
\end{eqnarray}
The configurations satisfying this equation are known as instantons, and they describe kinks in the otherwise constant configuration of the $\theta$ field. Their action consists of two parts:
\begin {enumerate}
\item 
The action cost of the kink, which scales as $2\pi r_0$, as the kink occurs only along the domain wall. It may be calculated for $\mathcal{J}=0$ with the assitance of the Euler-Lagrange equation, and results in
\begin{eqnarray}
S_0\approx 2.5\sqrt{\frac{uK}{2\pi}J}
\end{eqnarray}
\item
The action cost of the field being in a metastable minimum, which has an energy larger by $\Delta\epsilon=2\pi \frac{\mathcal{J}}{e}$. This contribution scales as $\pi r_0^2$, which is the size of the domain in which the field is in the higher energy state.
\end{enumerate}
Therefore, the total action which corresponds to a propagating domain wall in the presence of a nonvanishing current is
\begin{eqnarray}
S=2\pi r_0S_0-\pi^2r_0^2\frac{\mathcal{J}}{e}.
\end{eqnarray}
The minimum of this action occurs for $r_0=S_0/(\pi \frac{\mathcal{J}}{e})$, and corresponds to the total action
\begin{eqnarray}
\mathcal{S}=\frac{S_0^2e}{\mathcal{J}}.
\end{eqnarray}

However, this action is not the only contribution to the path integral that can be obtained from the saddle point approximation. Since the kink of the instanton is localized in time, it is feasible to assume that any number of instantons, seperated such that they can be considered as non interacting in the sense that the total action is simply the addition of $n$ single instanton actions $S_{\mbox{inst}}$, is also a saddle point configuration. The path integral that is obtained for the Green's function (A2) is therefore
\begin{eqnarray}
G(\tau)&\approx&\sum_n C^n\int_0^\tau\mbox{d}\tau_1\mbox{...}\int_0^{\tau_{n-1}}\mbox{d}\tau_n e^{-n\mathcal{S}}\notag\\
&=&\sum_n\frac{1}{n!}\left(C\tau e^{-\mathcal{S}}\right)^n\notag\\
&=&e^{C\tau e^{-\mathcal{S}}}
\end{eqnarray}
where the $n$-fold integration occurs due to space-time translation invariance; the value of $r$ at which the kink occurs can vary between $0$ and $r$ for $r\rightarrow\infty$. $C$ is a dimensionful constant which arises from the Gaussian fluctuations around the minima of action, and from normalization factors. 

Applying the analytic continuation, and relying on the fact that K must be pure imaginary \cite{Altland},
\begin{eqnarray}
G(t)\propto e^{-t|C|e^{-\mathcal{S}}},
\end{eqnarray}
and it follows that the tunneling rate from the metastable minimum is
\begin{eqnarray}
\Gamma=|C|e^{-\mathcal{S}}\propto e^{-\frac{\mathcal{S}^2e}{\mathcal{J}}}.
\end{eqnarray}

\bibliographystyle{apsrev}
\bibliography{Bibliography}

\begin{thebibliography}{31}
\expandafter\ifx\csname natexlab\endcsname\relax\def\natexlab#1{#1}\fi
\expandafter\ifx\csname bibnamefont\endcsname\relax
  \def\bibnamefont#1{#1}\fi
\expandafter\ifx\csname bibfnamefont\endcsname\relax
  \def\bibfnamefont#1{#1}\fi
\expandafter\ifx\csname citenamefont\endcsname\relax
  \def\citenamefont#1{#1}\fi
\expandafter\ifx\csname url\endcsname\relax
  \def\url#1{\texttt{#1}}\fi
\expandafter\ifx\csname urlprefix\endcsname\relax\def\urlprefix{URL }\fi
\providecommand{\bibinfo}[2]{#2}
\providecommand{\eprint}[2][]{\url{#2}}

\bibitem[{\citenamefont{Eisenstein and MacDonald}(2004)}]{MacDonald}
\bibinfo{author}{\bibfnamefont{J.~P.} \bibnamefont{Eisenstein}}
  \bibnamefont{and} \bibinfo{author}{\bibfnamefont{A.~H.}
  \bibnamefont{MacDonald}}, \bibinfo{journal}{Nature}
  \textbf{\bibinfo{volume}{432}}, \bibinfo{pages}{691} (\bibinfo{year}{2004}).

\bibitem[{\citenamefont{Seamons et~al.}(2007)\citenamefont{Seamons, Tibbetts,
  Reno, and Lilly}}]{Seamons2007}
\bibinfo{author}{\bibfnamefont{J.~A.} \bibnamefont{Seamons}},
  \bibinfo{author}{\bibfnamefont{D.~R.} \bibnamefont{Tibbetts}},
  \bibinfo{author}{\bibfnamefont{J.~L.} \bibnamefont{Reno}}, \bibnamefont{and}
  \bibinfo{author}{\bibfnamefont{M.~P.} \bibnamefont{Lilly}},
  \bibinfo{journal}{Applied Physics Letters} \textbf{\bibinfo{volume}{90}},
  \bibinfo{eid}{052103} (\bibinfo{year}{2007}),
  \urlprefix\url{http://scitation.aip.org/content/aip/journal/apl/90/5/10.1063/1.2437664}.

\bibitem[{\citenamefont{Nandi et~al.}(2012)\citenamefont{Nandi, Finck,
  Eisenstein, Pfeiffer, and West}}]{Nandi}
\bibinfo{author}{\bibfnamefont{D.}~\bibnamefont{Nandi}},
  \bibinfo{author}{\bibfnamefont{A.~D.~K.} \bibnamefont{Finck}},
  \bibinfo{author}{\bibfnamefont{J.~P.} \bibnamefont{Eisenstein}},
  \bibinfo{author}{\bibfnamefont{L.~N.} \bibnamefont{Pfeiffer}},
  \bibnamefont{and} \bibinfo{author}{\bibfnamefont{K.~W.} \bibnamefont{West}},
  \bibinfo{journal}{Nature} \textbf{\bibinfo{volume}{488}},
  \bibinfo{pages}{481} (\bibinfo{year}{2012}).

\bibitem[{\citenamefont{Gupta et~al.}(2011)\citenamefont{Gupta, Croxall, Waldie
  et~al.}}]{DasGupta2010-1}
\bibinfo{author}{\bibfnamefont{K.~D.} \bibnamefont{Gupta}},
  \bibinfo{author}{\bibfnamefont{A.~F.} \bibnamefont{Croxall}},
  \bibinfo{author}{\bibfnamefont{J.}~\bibnamefont{Waldie}},
  \bibnamefont{et~al.}, \bibinfo{journal}{Advances in Condesnsed Matter
  Physics} \textbf{\bibinfo{volume}{2011}} (\bibinfo{year}{2011}).

\bibitem[{\citenamefont{Laroche et~al.}(2014)\citenamefont{Laroche, Gervais,
  Lilly, and Reno}}]{Laroche}
\bibinfo{author}{\bibfnamefont{D.}~\bibnamefont{Laroche}},
  \bibinfo{author}{\bibfnamefont{G.}~\bibnamefont{Gervais}},
  \bibinfo{author}{\bibfnamefont{M.~P.} \bibnamefont{Lilly}}, \bibnamefont{and}
  \bibinfo{author}{\bibfnamefont{J.~L.} \bibnamefont{Reno}},
  \bibinfo{journal}{Science} \textbf{\bibinfo{volume}{343}},
  \bibinfo{pages}{631} (\bibinfo{year}{2014}),
  \eprint{http://www.sciencemag.org/content/343/6171/631.full.pdf},
  \urlprefix\url{http://www.sciencemag.org/content/343/6171/631.abstract}.

\bibitem[{\citenamefont{Spielman et~al.}(2000)\citenamefont{Spielman,
  Eisenstein, Pfeiffer, and West}}]{Spielman2000}
\bibinfo{author}{\bibfnamefont{I.~B.} \bibnamefont{Spielman}},
  \bibinfo{author}{\bibfnamefont{J.~P.} \bibnamefont{Eisenstein}},
  \bibinfo{author}{\bibfnamefont{L.~N.} \bibnamefont{Pfeiffer}},
  \bibnamefont{and} \bibinfo{author}{\bibfnamefont{K.~W.} \bibnamefont{West}},
  \bibinfo{journal}{Phys. Rev. Lett.} \textbf{\bibinfo{volume}{84}},
  \bibinfo{pages}{5808} (\bibinfo{year}{2000}),
  \urlprefix\url{http://link.aps.org/doi/10.1103/PhysRevLett.84.5808}.

\bibitem[{com()}]{comment-Min2008}
\bibinfo{note}{This is a one-dimensional analogue of the setup proposed by: H.
  Min, R. Bistritzer, S. Jung-Jung, and A. H. MacDonald, Phys. Rev. B
  \textbf{78}, 121401 (2008).}

\bibitem[{\citenamefont{Shelton and Tsvelik}(1996)}]{Tsvelik}
\bibinfo{author}{\bibfnamefont{D.~G.} \bibnamefont{Shelton}} \bibnamefont{and}
  \bibinfo{author}{\bibfnamefont{A.~M.} \bibnamefont{Tsvelik}},
  \bibinfo{journal}{Phys. Rev. B} \textbf{\bibinfo{volume}{53}},
  \bibinfo{pages}{14036} (\bibinfo{year}{1996}),
  \urlprefix\url{http://link.aps.org/doi/10.1103/PhysRevB.53.14036}.

\bibitem[{\citenamefont{Klesse and Stern}(2000)}]{Stern}
\bibinfo{author}{\bibfnamefont{R.}~\bibnamefont{Klesse}} \bibnamefont{and}
  \bibinfo{author}{\bibfnamefont{A.}~\bibnamefont{Stern}},
  \bibinfo{journal}{Phys. Rev. B} \textbf{\bibinfo{volume}{62}},
  \bibinfo{pages}{16912} (\bibinfo{year}{2000}),
  \urlprefix\url{http://link.aps.org/doi/10.1103/PhysRevB.62.16912}.

\bibitem[{\citenamefont{{Pustilnik} et~al.}(2003)\citenamefont{{Pustilnik},
  {Mishchenko}, {Glazman}, and {Andreev}}}]{Pustilnik}
\bibinfo{author}{\bibfnamefont{M.}~\bibnamefont{{Pustilnik}}},
  \bibinfo{author}{\bibfnamefont{E.~G.} \bibnamefont{{Mishchenko}}},
  \bibinfo{author}{\bibfnamefont{L.~I.} \bibnamefont{{Glazman}}},
  \bibnamefont{and} \bibinfo{author}{\bibfnamefont{A.~V.}
  \bibnamefont{{Andreev}}}, \bibinfo{journal}{Physical Review Letters}
  \textbf{\bibinfo{volume}{91}}, \bibinfo{eid}{126805} (\bibinfo{year}{2003}),
  \eprint{cond-mat/0208267}.

\bibitem[{\citenamefont{Giamarchi}(2003)}]{Giamarchi}
\bibinfo{author}{\bibfnamefont{T.}~\bibnamefont{Giamarchi}},
  \emph{\bibinfo{title}{Quantum Physics in One Dimension}}
  (\bibinfo{publisher}{Clarendon Press}, \bibinfo{year}{2003}).

\bibitem[{\citenamefont{Aristov et~al.}(2010)\citenamefont{Aristov, Dmitriev,
  Gornyi, Kachorovskii, Polyakov, and W\"olfle}}]{Aristov1}
\bibinfo{author}{\bibfnamefont{D.~N.} \bibnamefont{Aristov}},
  \bibinfo{author}{\bibfnamefont{A.~P.} \bibnamefont{Dmitriev}},
  \bibinfo{author}{\bibfnamefont{I.~V.} \bibnamefont{Gornyi}},
  \bibinfo{author}{\bibfnamefont{V.~Y.} \bibnamefont{Kachorovskii}},
  \bibinfo{author}{\bibfnamefont{D.~G.} \bibnamefont{Polyakov}},
  \bibnamefont{and} \bibinfo{author}{\bibfnamefont{P.}~\bibnamefont{W\"olfle}},
  \bibinfo{journal}{Phys. Rev. Lett.} \textbf{\bibinfo{volume}{105}},
  \bibinfo{pages}{266404} (\bibinfo{year}{2010}),
  \urlprefix\url{http://link.aps.org/doi/10.1103/PhysRevLett.105.266404}.

\bibitem[{\citenamefont{Bockrath et~al.}(1999)\citenamefont{Bockrath, Cobden,
  Lu, Rinzler, Smalley, Balents, and McEuen}}]{Bockarth}
\bibinfo{author}{\bibfnamefont{M.}~\bibnamefont{Bockrath}},
  \bibinfo{author}{\bibfnamefont{D.~H.} \bibnamefont{Cobden}},
  \bibinfo{author}{\bibfnamefont{J.}~\bibnamefont{Lu}},
  \bibinfo{author}{\bibfnamefont{A.~G.} \bibnamefont{Rinzler}},
  \bibinfo{author}{\bibfnamefont{R.~E.} \bibnamefont{Smalley}},
  \bibinfo{author}{\bibfnamefont{L.}~\bibnamefont{Balents}}, \bibnamefont{and}
  \bibinfo{author}{\bibfnamefont{P.~L.} \bibnamefont{McEuen}},
  \bibinfo{journal}{Nature} \textbf{\bibinfo{volume}{397}},
  \bibinfo{pages}{598} (\bibinfo{year}{1999}).

\bibitem[{\citenamefont{Auslaender et~al.}(2002)\citenamefont{Auslaender,
  Yacoby, de~Picciotto, Baldwin, Pfeiffer, and West}}]{Auslaender}
\bibinfo{author}{\bibfnamefont{O.~M.} \bibnamefont{Auslaender}},
  \bibinfo{author}{\bibfnamefont{A.}~\bibnamefont{Yacoby}},
  \bibinfo{author}{\bibfnamefont{R.}~\bibnamefont{de~Picciotto}},
  \bibinfo{author}{\bibfnamefont{K.~W.} \bibnamefont{Baldwin}},
  \bibinfo{author}{\bibfnamefont{L.~N.} \bibnamefont{Pfeiffer}},
  \bibnamefont{and} \bibinfo{author}{\bibfnamefont{K.~W.} \bibnamefont{West}},
  \bibinfo{journal}{Science} \textbf{\bibinfo{volume}{295}},
  \bibinfo{pages}{825} (\bibinfo{year}{2002}),
  \eprint{http://www.sciencemag.org/content/295/5556/825.full.pdf},
  \urlprefix\url{http://www.sciencemag.org/content/295/5556/825.abstract}.

\bibitem[{\citenamefont{Jompol et~al.}(2009)\citenamefont{Jompol, Ford,
  Griffiths, Farrer, Jones, Anderson, Ritchie, Silk, and Schofield}}]{Jompol}
\bibinfo{author}{\bibfnamefont{Y.}~\bibnamefont{Jompol}},
  \bibinfo{author}{\bibfnamefont{C.~J.~B.} \bibnamefont{Ford}},
  \bibinfo{author}{\bibfnamefont{J.~P.} \bibnamefont{Griffiths}},
  \bibinfo{author}{\bibfnamefont{I.}~\bibnamefont{Farrer}},
  \bibinfo{author}{\bibfnamefont{G.~A.~C.} \bibnamefont{Jones}},
  \bibinfo{author}{\bibfnamefont{D.}~\bibnamefont{Anderson}},
  \bibinfo{author}{\bibfnamefont{D.~A.} \bibnamefont{Ritchie}},
  \bibinfo{author}{\bibfnamefont{T.~W.} \bibnamefont{Silk}}, \bibnamefont{and}
  \bibinfo{author}{\bibfnamefont{A.~J.} \bibnamefont{Schofield}},
  \bibinfo{journal}{Science} \textbf{\bibinfo{volume}{325}},
  \bibinfo{pages}{597} (\bibinfo{year}{2009}),
  \eprint{http://www.sciencemag.org/content/325/5940/597.full.pdf},
  \urlprefix\url{http://www.sciencemag.org/content/325/5940/597.abstract}.

\bibitem[{\citenamefont{Hyart and Rosenow}(2011)}]{Rosenow}
\bibinfo{author}{\bibfnamefont{T.}~\bibnamefont{Hyart}} \bibnamefont{and}
  \bibinfo{author}{\bibfnamefont{B.}~\bibnamefont{Rosenow}},
  \bibinfo{journal}{Phys. Rev. B} \textbf{\bibinfo{volume}{83}},
  \bibinfo{pages}{155315} (\bibinfo{year}{2011}),
  \urlprefix\url{http://link.aps.org/doi/10.1103/PhysRevB.83.155315}.

\bibitem[{\citenamefont{Mahan}(2000)}]{Mahan}
\bibinfo{author}{\bibfnamefont{G.~D.} \bibnamefont{Mahan}},
  \emph{\bibinfo{title}{Many-Particle Physics}} (\bibinfo{publisher}{Springer
  Science}, \bibinfo{year}{2000}).

\bibitem[{\citenamefont{Tinkham}(2004)}]{Tinkham}
\bibinfo{author}{\bibfnamefont{M.}~\bibnamefont{Tinkham}},
  \emph{\bibinfo{title}{Introduction to Superconductivity}}, Dover books on
  physics and chemistry (\bibinfo{publisher}{Dover Publications},
  \bibinfo{year}{2004}), ISBN \bibinfo{isbn}{9780486435039},
  \urlprefix\url{http://books.google.co.il/books?id=k6AO9nRYbioC}.

\bibitem[{\citenamefont{White et~al.}(2002)\citenamefont{White, Affleck, and
  Scalapino}}]{White2002}
\bibinfo{author}{\bibfnamefont{S.~R.} \bibnamefont{White}},
  \bibinfo{author}{\bibfnamefont{I.}~\bibnamefont{Affleck}}, \bibnamefont{and}
  \bibinfo{author}{\bibfnamefont{D.~J.} \bibnamefont{Scalapino}},
  \bibinfo{journal}{Phys. Rev. B} \textbf{\bibinfo{volume}{65}},
  \bibinfo{pages}{165122} (\bibinfo{year}{2002}),
  \urlprefix\url{http://link.aps.org/doi/10.1103/PhysRevB.65.165122}.

\bibitem[{\citenamefont{Pereira and Sela}(2010)}]{Pereira}
\bibinfo{author}{\bibfnamefont{R.~G.} \bibnamefont{Pereira}} \bibnamefont{and}
  \bibinfo{author}{\bibfnamefont{E.}~\bibnamefont{Sela}},
  \bibinfo{journal}{Phys. Rev. B} \textbf{\bibinfo{volume}{82}},
  \bibinfo{pages}{115324} (\bibinfo{year}{2010}),
  \urlprefix\url{http://link.aps.org/doi/10.1103/PhysRevB.82.115324}.

\bibitem[{\citenamefont{Teber}(2007)}]{Teber}
\bibinfo{author}{\bibfnamefont{S.}~\bibnamefont{Teber}},
  \bibinfo{journal}{Phys. Rev. B} \textbf{\bibinfo{volume}{76}},
  \bibinfo{pages}{045309} (\bibinfo{year}{2007}),
  \urlprefix\url{http://link.aps.org/doi/10.1103/PhysRevB.76.045309}.

\bibitem[{\citenamefont{Flensberg}(1998)}]{Flensberg}
\bibinfo{author}{\bibfnamefont{K.}~\bibnamefont{Flensberg}},
  \bibinfo{journal}{Phys. Rev. Lett.} \textbf{\bibinfo{volume}{81}},
  \bibinfo{pages}{184} (\bibinfo{year}{1998}),
  \urlprefix\url{http://link.aps.org/doi/10.1103/PhysRevLett.81.184}.

\bibitem[{\citenamefont{Nazarov and Averin}(1998)}]{Nazarov1}
\bibinfo{author}{\bibfnamefont{Y.~V.} \bibnamefont{Nazarov}} \bibnamefont{and}
  \bibinfo{author}{\bibfnamefont{D.~V.} \bibnamefont{Averin}},
  \bibinfo{journal}{Phys. Rev. Lett.} \textbf{\bibinfo{volume}{81}},
  \bibinfo{pages}{653} (\bibinfo{year}{1998}),
  \urlprefix\url{http://link.aps.org/doi/10.1103/PhysRevLett.81.653}.

\bibitem[{\citenamefont{Ponomarenko and Averin}(2000)}]{Ponomarenko}
\bibinfo{author}{\bibfnamefont{V.~V.} \bibnamefont{Ponomarenko}}
  \bibnamefont{and} \bibinfo{author}{\bibfnamefont{D.~V.}
  \bibnamefont{Averin}}, \bibinfo{journal}{Phys. Rev. Lett.}
  \textbf{\bibinfo{volume}{85}}, \bibinfo{pages}{4928} (\bibinfo{year}{2000}),
  \urlprefix\url{http://link.aps.org/doi/10.1103/PhysRevLett.85.4928}.

\bibitem[{\citenamefont{Fiete et~al.}(2006)\citenamefont{Fiete, Le~Hur, and
  Balents}}]{Fiete}
\bibinfo{author}{\bibfnamefont{G.~A.} \bibnamefont{Fiete}},
  \bibinfo{author}{\bibfnamefont{K.}~\bibnamefont{Le~Hur}}, \bibnamefont{and}
  \bibinfo{author}{\bibfnamefont{L.}~\bibnamefont{Balents}},
  \bibinfo{journal}{Phys. Rev. B} \textbf{\bibinfo{volume}{73}},
  \bibinfo{pages}{165104} (\bibinfo{year}{2006}),
  \urlprefix\url{http://link.aps.org/doi/10.1103/PhysRevB.73.165104}.

\bibitem[{\citenamefont{Dresselhaus et~al.}(2001)\citenamefont{Dresselhaus,
  Dresselhaus, and Avouris}}]{Dresselhaus}
\bibinfo{author}{\bibfnamefont{S.}~\bibnamefont{Dresselhaus}},
  \bibinfo{author}{\bibfnamefont{G.}~\bibnamefont{Dresselhaus}},
  \bibnamefont{and} \bibinfo{author}{\bibfnamefont{P.}~\bibnamefont{Avouris}},
  \emph{\bibinfo{title}{Carbon Nanotubes: Synthesis, Structure, Properties, and
  Applications}}, Physics and astronomy online library
  (\bibinfo{publisher}{Springer}, \bibinfo{year}{2001}), ISBN
  \bibinfo{isbn}{9783540410867},
  \urlprefix\url{http://books.google.co.il/books?id=dkvDhZJnafgC}.

\bibitem[{\citenamefont{Kuemmeth et~al.}(2008)\citenamefont{Kuemmeth, Ilani,
  Ralph, and McEuen}}]{Ilani}
\bibinfo{author}{\bibfnamefont{F.}~\bibnamefont{Kuemmeth}},
  \bibinfo{author}{\bibfnamefont{S.}~\bibnamefont{Ilani}},
  \bibinfo{author}{\bibfnamefont{D.~C.} \bibnamefont{Ralph}}, \bibnamefont{and}
  \bibinfo{author}{\bibfnamefont{P.~L.} \bibnamefont{McEuen}},
  \bibinfo{journal}{Nature} \textbf{\bibinfo{volume}{452}},
  \bibinfo{pages}{448} (\bibinfo{year}{2008}), ISSN \bibinfo{issn}{7186},
  \urlprefix\url{http://dx.doi.org/10.1038/nature06822}.

\bibitem[{\citenamefont{Kane et~al.}(1997)\citenamefont{Kane, Balents, and
  Fisher}}]{Kane1997}
\bibinfo{author}{\bibfnamefont{C.}~\bibnamefont{Kane}},
  \bibinfo{author}{\bibfnamefont{L.}~\bibnamefont{Balents}}, \bibnamefont{and}
  \bibinfo{author}{\bibfnamefont{M.~P.~A.} \bibnamefont{Fisher}},
  \bibinfo{journal}{Phys. Rev. Lett.} \textbf{\bibinfo{volume}{79}},
  \bibinfo{pages}{5086} (\bibinfo{year}{1997}),
  \urlprefix\url{http://link.aps.org/doi/10.1103/PhysRevLett.79.5086}.

\bibitem[{\citenamefont{Yao et~al.}(1999)\citenamefont{Yao, Postma, Balents,
  and Dekker}}]{Yao1999}
\bibinfo{author}{\bibfnamefont{Z.}~\bibnamefont{Yao}},
  \bibinfo{author}{\bibfnamefont{H.~W.~C.} \bibnamefont{Postma}},
  \bibinfo{author}{\bibfnamefont{L.}~\bibnamefont{Balents}}, \bibnamefont{and}
  \bibinfo{author}{\bibfnamefont{C.}~\bibnamefont{Dekker}},
  \bibinfo{journal}{Nature} \textbf{\bibinfo{volume}{402}},
  \bibinfo{pages}{273} (\bibinfo{year}{1999}).

\bibitem[{\citenamefont{von Delft and Schoeller}(1998)}]{Delft}
\bibinfo{author}{\bibfnamefont{J.}~\bibnamefont{von Delft}} \bibnamefont{and}
  \bibinfo{author}{\bibfnamefont{H.}~\bibnamefont{Schoeller}},
  \bibinfo{journal}{Annalen der Physik} \textbf{\bibinfo{volume}{7}},
  \bibinfo{pages}{225} (\bibinfo{year}{1998}), ISSN \bibinfo{issn}{1521-3889},
  \urlprefix\url{http://dx.doi.org/10.1002/(SICI)1521-3889(199811)7:4<225::AID-ANDP225>3.0.CO;2-L}.

\bibitem[{\citenamefont{Altland and Simons}(2010)}]{Altland}
\bibinfo{author}{\bibfnamefont{A.}~\bibnamefont{Altland}} \bibnamefont{and}
  \bibinfo{author}{\bibfnamefont{B.}~\bibnamefont{Simons}},
  \emph{\bibinfo{title}{Condensed Matter Field Theory}}
  (\bibinfo{publisher}{Cambridge University Press}, \bibinfo{year}{2010}).

\end{thebibliography}

\end{document}